# Optically and elastically assembled plasmonic nanoantennae for spatially resolved characterization of chemical composition in soft matter systems using surface enhanced spontaneous and stimulated Raman scattering

Haridas Mundoor,[1] Taewoo Lee,[1] Derek G. Gann,[1] Paul J Ackerman[1], Bohdan Senyuk,[1] Jao van de Lagemaat,[2,3a)] Ivan I. Smalyukh[1,2,4a)]

*[1]Department of Physics and Liquid Crystal Materials Research Center, University of Colorado, Boulder, Colorado 80309, USA.*
*[2]Renewable and Sustainable Energy Institute, National Renewable Energy Laboratory and University of Colorado, Boulder, Colorado 80309, USA.*
*[3]National Renewable Energy Laboratory, Golden, Colorado 80401, USA.*
*[4]Department of Electrical, Computer and Energy Engineering and Materials Science Engineering Program, University of Colorado, Boulder, Colorado 80309, USA.*

We present a method to locally probe spatially varying chemical composition of soft matter systems by use of optically controlled and elastically self-assembled plasmonic nanoantennae. Disc-shaped metal particles with sharp irregular edges are optically trapped, manipulated, and assembled into small clusters to provide a strong enhancement of the Raman scattering signal coming from the sample regions around and in-between these particles. As the particles are reassembled and spatially translated by computer-controlled laser tweezers, we probe chemical composition as a function of spatial coordinates. This allows us to reliably detect tiny quantities of organic molecules, such as capping ligands present on various nanoparticles, as well as to probe chemical composition of the interior of liquid crystal defect cores that can be filled with, for example, polymer chains. The strong electromagnetic field enhancement of optically-manipulated nanoparticles' rough surfaces is demonstrated in different forms of spectroscopy and microscopy, including enhanced spontaneous Raman scattering, coherent anti-Stokes Raman scattering, and stimulated Raman scattering imaging modes.

[a)]Electronic mail: Jao.vandeLagemaat@nrel.gov, ivan.smalyukh@colorado.edu

## I. INTRODUCTION

Surface-enhanced Raman scattering (SERS) spectroscopy[1-10] is a powerful tool for chemical identification and has been used extensively for imaging of biological tissues[10]. Earlier experiments on SERS were performed with metal-coated substrates with suitably tuned roughness, as needed for enhancement of the Raman Scattering signal. However, with the advent of nanoparticles and considerable progress in their wet chemical synthesis, metal nano structures of different shape and size were made possible.[11] Such metal nanoparticles have been frequently used for SERS.[6-9] Nano structures of noble metals like gold and silver are especially suitable for SERS applications because of their tunable optical responses in the visible region, through surface plasmon resonance (SPR) effects. Raman enhancement of several orders of magnitude can be obtained by suitably arranging metal particles in dimers, trimers and so on,[5-7] as well as by using particles with sharp edges.[9] However, manipulating such particles in different media and controlling the position within the sample where the enhancement occurs has been a challenge, although certain metal nanoparticles can be manipulated with optical tweezers and can be localized in topological defects in liquid crystalline (LC) host media.[12,13]

In this work, we combine SERS with optical manipulation and LC-mediated self-assembly of metal nanoparticles to demonstrate chemical identification of molecules and individual nanoparticles with the help of surface-enhanced Raman techniques, including Raman micro-spectroscopy, coherent anti-Stokes Raman scattering (CARS), and stimulated Raman scattering (SRS) polarizing microscopies (PM).

Disc-shaped gold nanoparticles with sharp, irregular edges, which act as plasmonic nanoantennae, are used to enhance Raman signals in both imaging and spectroscopy. We probe chemical composition using the SERS signal from different selected regions controlled through a nanometer precision stage and optical trapping system. The disc shaped particles with irregular sharp edges are effectively manipulated by laser tweezers and assembled into various nanostructures using the elastic distortions of the LC host medium. We demonstrate that optically and elastically assembled, plasmonic nanoantennae enable enhancement of Raman signal from LC molecules (on average, ~500 times), the capping ligands of different nanoparticles, and from polymer chains segregating into topological defect lines.

## II. EXPERIMENTAL

### A. Experimental setup of Raman micro-spectrometer with optical trapping

The experimental setup [Fig. 1] consists of a Raman micro-spectrometer and an optical trapping system integrated with an inverted microscope (IX 71, Olympus), equipped with a computer-controlled motorized stage with nanometer precision by the homemade stage control software. The Raman micro-spectrometer setup was implemented using SpectraPro-275 spectrometer (Acton Research Corporation), an electron multiplying charge coupled device (EMCCD, iXon3 888, Andor Technology), and a continuous-wave 632.8 nm laser, which was selected to minimize the absorption by gold nanoparticles. The laser beam is sent through a dichroic mirror, guided to the microscope with suitable mirrors and focused on the sample using a 100x oil immersion objective with adjustable numerical aperture (NA) of 0.6 -1.3. The Raman-Stokes signals from the sample were separated from the Rayleigh line using a long pass, edge filter before being sent to a spectrometer equipped with a grating of 600g/mm and finally recorded the spectra with an EMCCD detector. Laser trapping utilized 1064 nm laser light (guided to the microscope objective with a dichroic mirror) and could be based on a single stationary, focused laser beam and computer-controlled microscope stage or a holographic optical trapping setup (used in conjunction with CARS-PM and SRS-PM) described elsewhere[12].

Optical trapping of gold nanoparticles was characterized in dark field imaging mode using a dark-field condenser with NA = 1.2–1.4 (U-DCW) and an objective with NA = 0.6 (both from Olympus). The power of the trapping beam was measured just before entering the objective. The Laguerre-Gaussian (LG) beam was generated with a phase hologram placed in the laser beam path obtained with a two-dimensional LC-based spatial light modulator (S.L.M, Boulder Nonlinear Systems). The integrated manipulation and imaging/spectroscopy system is capable of trapping nanometer sized particles in LC media, moving them through the medium using a motorized stage with nanometer precision to assemble desired nano structures, and simultaneously collecting the Raman spectra from such assemblies with the 632.8 nm excitation laser beam focused congruently on the sample.

Coherent Raman images of metal nanoparticle assemblies in LC medium were obtained by using SRS-PM and CARS-PM [Fig 1(b)][14,15]. A single Ti:Sapphire oscillator (Chameleon Ultra II, Coherent, 140 fs, 80 MHz) provides a pump pulse and also generates a Stokes pulse by use of a photonic crystal fiber (FemtoWhite-800, NKT Photonics), which is then modulated at 4 MHz by using an electro-optic modulator (M350-160 KD*P, Conoptics). The pump and Stokes beams are recombined spatially and temporally before directing them into an XY-mirror scanning unit (FV300, Olympus) and an inverted microscope (IX81, Olympus). The enhanced SRS signal is obtained from demodulation of the pump beam, i.e. stimulated Raman loss, using a photodiode and a lock-in-amplifier (SR844, Stanford Research Systems). The surface-enhanced CARS signal is simultaneously collected by a photomultiplier tube in forward detection mode; all signals are read by a data acquisition system and finally processed by a computer software to form images.

### B. Sample preparation

Short (20×40nm) gold nanorods (SR) with capping ligands polystyrene (PS) or methoxy poly(ethylene glycol) thiol (mPEG) and disc-shaped gold nanoparticles with sharp irregular edges called 'Nanobursts' (NB)[12,13] were obtained from Nanopartz Inc. The relatively poly-disperse NSOL-functionalized gold NB nanoparticles had average lateral size ~ 500 nm, as indicated by the TEM images shown in Fig 2(a). The random sharp edges of NB particles allow them to efficiently localize the electromagnetic field and thereby provide high field enhancement for the incident electromagnetic wave [Fig 2(b)]. When dispersed in LCs, the NB particles possessed homeotropic anchoring of LC molecules at their surfaces, which led to quadrupolar elastic distortions [12,13] of director field $\boldsymbol{n}(\boldsymbol{r})$ with a half-integer disclination around particle's perimeter (inset in Fig. 1). These NB nanoplatelets spontaneously align normal to the far-field director $\boldsymbol{n_0}$ (inset in Fig. 1). They can be trapped effectively in a LC medium and multiple particles can be assembled into colloidal superstructures using an optical trapping system, as shown in the optical micrographs in Fig 2. (d)-(g).

Our samples were prepared by dispersing the gold nanoparticles in an aligned 4-Cyano-4′-pentylbiphenyl (5CB) LC. Glass cells were prepared with a gap spacing of 3 or 10 μm and then filled with LCs containing nanoparticles using capillary forces. The glass substrates used for preparing cells with homeotropic far-field alignment of the LC molecules were treated with Dimethyl octadecyl [3-(trimethoxysilyl) propyl] ammonium chloride (DMOAP). The glass substrates used for fabricating the cells with planar boundary conditions were spin-coated with polyimide PI2555 (HD MicroSystem) at 3000 rpm for 30 sec, baked (5 min at 90˚C followed by 1 h at 180˚C) and unidirectionally rubbed. For the experiments on partially polymerized samples, a photo-polymerizable LC mixture was prepared by mixing 5CB, diacrylate nematic reactive mesogen (RM), and ~1% of photo-initiator (Irgacure 184). A suitable small amount (< 0.1wt %) of NB particles was dispersed in the polymerizeable LC mixture before filling in the cell.

### C. Simulation of electromagnetic field enhancement of NB particles

The electromagnetic field enhancements of NB particles in LC medium were calculated using the finite element method, with the help of commercially available software (COMSOL Multiphysics). The irregular sharp edges of the NB particles were reproduced in the calculation by rendering it from the TEM images shown in Fig. 2(a). Since the NB platelets spontaneously orient orthogonally to the far-field director $\boldsymbol{n_0}$ in a homeotropic LC cell (inset of Fig. 1(a)), the effective refractive index of the medium is equal to the ordinary refractive index for all polarizations of light incident on the cell along $\boldsymbol{n_0}$. The dielectric constant of the matrix for the two simulated electric energy density distributions shown in Fig. 2(b,c) was set equal to that of square of the ordinary refractive index of 5CB (2.34 ). The incident electromagnetic wave (632.8 nm) passes through the computational volume from above (Z direction) along the direction of $\boldsymbol{n_0}$. The simulation volume was defined as a cylindrical box with diameter 1.5 μm and height 1 μm, with the NB structure located at the center of the box, enclosed between perfectly matched layers in all directions.

## III. RESULTS

### A. Optical trapping of gold NB particles

Certain types of gold nanoparticles can be effectively trapped and manipulated using optical tweezers[16-18]. We have observed such a stable trapping of NB particles dispersed in nematic LCs [Fig. 3(a)]. We used low laser powers to avoid heating of gold nanoparticles[18] and local melting of the surrounding LC. Lateral displacements of trapped NBs within a planar cell exhibited dependence on the polarization of the laser beam with respect to the director $\boldsymbol{n_0}$ and the NB orientation [Fig. 3(a)]. This is natural as the trapping potential and forces are expected to be highly anisotropic and polarization dependent because of the polarization dependence of the refractive index of the LC host and particle's shape anisotropy. The direction-averaged trapping stiffness of optical traps in a homeotropic cell increases linearly with the laser power at relatively low powers [Fig. 3(b)]. The laser trapping forces depend on the detailed shape and orientation of plasmonic NB nanoparticles, but always have order of magnitude values similar to what we present in Fig. 3b. The ratio between scattering and gradient forces are known to significantly vary as the scattering and absorption cross-sections change depending on the orientation of NB with respect to polarization of the trapping beam[16]. In the LC solvent, additional contribution to trapping forces comes from the contrast of refractive index between uniform alignment of the far field and the LC distortions around NB particles, as well as the dependence of the effective refractive index distribution on the trapping beam's polarization[19]. Consequently, the average trap stiffness at a laser power of $P$ = 32 mW in a planar cell (~ 3.5 pN/μm) was found to be different from that in a homeotropic LC cell (~ 1.7 pN/μm). Additionally, the NB nanoparticle can be elastically attracted and trapped, as well as manipulated by use of local melting of the LC. In addition to the optical trapping of NBs, we observed a transfer of the angular momentum from the vortex LG beam to the nanoparticle resembling that discussed in earlier literature reports[20]. The NBs rotate around the beam's propagation direction [Fig. 3(c)]. The direction of rotation depends on the topological charge $l$ of the beam, being clockwise for negative $l$ and counterclockwise for positive $l$ [Fig. 3(c, d)]. The rate also shows dependence on laser power and $l$ [Fig. 3(d)]. The rotation rate increases with increasing laser power and with decreasing the integer $|l| > 0$ and is, for example, ~ 0.5 Hz at $P$ = 40 mW and $l = \pm 2$ [Fig. 3(d)]. The angular velocity is dependent on the angle between the line connecting the centers of the LG beam, NP and ***n*** [Fig. 3(c)][21] . This dependence is caused by the anisotropy of LC's viscous drag forces as well as by the complex dependence of laser trapping forces on the particle's position in such an experiment. The controlled trapping and dynamics of NB particles allow us to effectively use them as movable nanoantennae for the enhancement of weak spontaneous and stimulated Raman scattering signals, as discussed below.

**B. Enhancement of Raman signals of 5CB by assemblies of NB particles**

The ability to optically trap NB nanoparticles allows us to move them through the studied medium either through dynamically changing holograms displayed by the SLM or with the help of a motorized stage spatially translating the sample. Since the near contact-assembly of NBs yields even stronger field enhancement in the inter-particle regions, we can optically organize multiple particles into side-by-side assemblies, which in LCs are then held together by elastic forces, as shown in Fig. 2(d-g). The electromagnetic field simulation for assemblies of NB particles [Fig.2(b,c)] shows a strong electromagnetic field enhancement at the sharp edges of the particles, especially in the inter-particle region. The higher field enhancement at the adjacent edges of the particles and in the region between them is due to the coupled electromagnetic field emerging from the dipole and quadrupole interactions between the particles. Interestingly, the electromagnetic enhancement depends strongly on the used linear polarization of the incident electric field [Fig. 2(b,c)].

The Raman scattering signal from 5CB molecules located in such a 'near-field' region of the particles is strongly enhanced. Figure 4 shows Raman spectra collected from a region of a homeotropically aligned 5CB cell with the NB particle. To calculate the enhancement factor, a reference spectrum of 5CB [Fig. 4(a)] was recorded from the same sample, moving the beam away from the NB particles. When calculating the enhancement factor, to mitigate the influence of temporal fluctuations in the intensity due to positional and orientational fluctuations of the particles on the enhancement, each Raman spectrum was collected over a relatively long integration time of 0.5s and then 1000 of such spectra were averaged using the EMCCD software. It is clearly evident from the comparison of Figs. 4(a) and 4(b) that the Raman lines corresponding to 5CB are enhanced considerably in the presence of NB. A close inspection of Fig. 4(b) reveals additional lines which do not correspond to 5CB. In order to investigate the origin of these additional lines, a Raman spectrum was collected from NB particles spin coated on a glass substrate, Fig. 4(c), which indicates that these additional lines originate from the NSOL capping ligands of NBs. As compared to regular 5CB spectra, the average enhancement factor of Raman signal from 5CB molecules in the presence of NB particles is ~500, although the values varies from 200 -700 for different NB particles. The calculations of this factor were performed by taking the ratios of integrated intensities of SERS signal corresponding to the characteristic line at 1158 cm $^{-1}$ collected from the NB location and Raman signal collected after moving the beam away from NB particles in LC medium, normalized with respect to the number of 5CB molecules in the excitation volume, as commonly done in SERS literature and described in more details elsewhere[7]. The achieved enhancement factor is somewhat limited by the fact that the Raman spectra measured from the NB locations are convoluted because of the larger volume of the excitation beam as compared to NB assembly size, which is only about several hundreds of nanometer along the light propagation direction. The electromagnetic field enhancement is limited to a sample region very close to the NB particles and decays rapidly with the distance [Fig. 2 (b,c)]. Consequently, Raman signals from 5CB molecules that are farther away from the NB particles are practically unaffected by the field enhancement, adding up together with the maximally SERS enhanced signal of 5CB molecules near NB and yielding the Raman spectra shown in Fig 4(b). The enhancement factor calculated from the simulations was found to be dependent on the polarization of the incident electromagnetic wave, whether it is parallel or perpendicular to the line joining center of the particles. The calculated values of enhancement factor based on the simulated values corresponding to these two polarization were 115 and 194. The higher values of the enhancement factors observed experimentally can be explained by such factors as (1) the waist of the beam tightly focused by the high-NA objective is slightly smaller than the diameter of the volume used in simulations, (2) the local director field close to the NB particles [Fig. (1)] is distorted and complex, having a noticeable in-plane component, which may result in an additional increase of the intensity of the Raman signal, (3) each particle has its own unique roughness features which cannot be fully accounted for in experiments but can substantially influence the enhancement factor, especially small features that we cannot resolve in electron microscopy micrographs, and finally (4), the simulations only probe a single relative orientation and distance of the two particles.

### C. Surface enhanced coherent Raman imaging in SRS and CARS microscopy modes

The results presented in Fig. 4 indicate that there is a strong enhancement of the conventional spontaneous Raman signal from 5CB molecules in presence of NB particles. Using NB particles dispersed in 5CB, we have also performed nonlinear Raman scattering measurements in the SRS-PM and CARS-PM imaging modes [Fig. 5]. The bright regions around the NB particles in SRS [Fig. 5(a)] and CARS [Fig. 5(b)] images represent the enhanced nonlinear stimulated Raman signals from the C-N stretching vibration mode (2228 $cm^{-1}$) of 5CB molecules around the NB particles. The vertical cross-sectional image shown in Fig. 5(a) demonstrates submicron spatial resolution along the microscope’s optical axis and also indicates that the enhancement of the Raman scattering occurs on the periphery of the NB particles. Although it is known that SRS and CARS nonlinear processes allow for obtaining stimulated Raman signal about a million times stronger as compared to spontaneous Raman scattering, these signals are still relatively weak when collected in the scanning microscopy experimental setting at used pixel dwell times of 0.2 ms, at which no signal beyond background can be detected from the LC sample regions far away from the particles. At the same scanning rate, both SRS and CARS signals are strong and exhibit different dependencies on polarization of laser excitation light [Fig. 5]. These results indicate that the NB particles can also further enhance stimulated Raman signals originating SRS and CARS nonlinear optical processes. The different polarization dependencies of SRS and CARS signals around the NB particles can be explained by a combination of two effects: (1) the electromagnetic field enhancement around particles is polarization dependent and also varies as a function of the detailed structure of the irregular edges around the NB [Fig. 2(b,c)]; (2) for polarizer-free detection, the SRS ($\propto \cos^4\beta$) and CARS ($\propto \cos^6\beta$) signals are dependent on the angle $\beta$ between the collinear polarizations of laser excitation light and locally distorted LC director $\boldsymbol{n}$, which forms a complex three-dimensional configuration and exhibits a singular defect line (inset of Fig. 1).

### D. Gold NB particles as probes of chemical inhomogeneity in composite systems

The presence of a signal from NSOL capping ligands in the Raman spectra collected from NB particles dispersed in 5CB [Fig. 4(b)] and when spin coated on a glass substrate [Fig. 4(c)] indicates a strong enhancement of the electromagnetic field sufficient for detection of very small amount of capping ligands. This result points to the potential of using such particles to probe chemical composition in composite systems, such as capping ligands of other nanoparticles or different dopants. To show this, we have co-dispersed NB and SR particles with different capping ligands mPEG and PS and measured the Raman spectra after localizing SR in the region between two NBs. Separate 5CB cells were prepared with dispersed NB and SR particles with the capping ligands mPEG and PS. In our experiments, The SR particles were detected using a dark field microscopy imaging mode and then the NB were moved close to the SR with the help of the optical trapping system, forming assemblies of particles as shown in the inset of Fig. 6. The enhanced Raman spectra from assemblies of short nanorods and NB particles, as shown in Fig. 6(b) and (c), are remarkably different from the spectra of NB particles in 5CB [Fig. 6 (a)]. The characteristic lines of the respective capping ligands (mPEG and PS) of the nanorods are clearly visible, along with the Raman lines from 5CB (indicated by star symbol) and NSOL. Although some of the characteristic lines of mPEG and PS (dashed lines in Fig. 6(b) and (c)) overlap with that of NSOL and 5CB, a comparison of the peak intensities of these lines indicates that we observe a strong Raman signal from the capping ligands of the SR particles. It is clear that the strong electromagnetic field due to NB particles, together with that from SR particle, enhances the Raman scattering signals from the capping ligands of NB and the nanorods, as needed for their identification.

We have used this approach to inspect the variation of polymer concentration in a partially polymerized LC cell through the use of the intensity of Raman signal from polymer as a measure of their relative concentration. A twisted nematic cell with 5CB, RM and NB was prepared by following the method described above. The samples were inspected using dark field microscopy to probe locations of NB particles inside the defects and, if needed, they were moved around using optical trapping before the photo-polymerization process. Raman measurements of the sample before polymerization did not indicate concentration gradients that would be strong enough to be detected, which is consistent with the good molecular-level mixing of the RM LC monomer and small-molecule LC. To inspect the variation of polymer concentration in the polymerized LC, we located a region with a NB particle trapped inside a defect. The Raman signal from around NB particles inside defects was then compared with the Raman signal from the adjacent regions in the bulk [Fig. 7]. Figures 7 (b) and 7(c) represent the spectra collected from NB located in the bulk and within defects in a polymerized LC sample, respectively. The spectra show the characteristic Raman lines from the polymer with different relative Raman intensities between 5CB and polymer chains (refer to dark diamond and star symbols) as depicted in Fig. 7(a). Higher intensities of the Raman lines of the polymer chains in Fig. 7(c) compared to Fig. 7(b) reveals higher concentrations of polymer chains inside LC defects as compared to the bulk regions away from defects. This indicates that the polymer chains segregate into the isotropic regions of the LC defects during the photo-polymerization process, which is natural as they represent a lower cost of free energy when localized within defects as compared to the uniformly aligned LC host in which they induce energetically costly elastic distortions. This result shows that plasmonic metal nanoparticles are suitable for nanoantennae-like probes with the potential applications in exploiting nanoscale composition of soft matter composites.

## IV. CONCLUSIONS.

We have developed an experimental tool for spatially resolved characterization of chemical composition in soft matter systems by using surface-enhanced spontaneous and stimulated Raman scattering with simultaneous optical manipulation of particles with suitable nanometer-sized structures. These structures produce large surface electromagnetic field enhancements and, as such, can act as optically and elastically assembled plasmonic nanoantennae for probing chemical composition. The optical trapping enables us to manipulate these nanoparticles through a soft matter medium with nanometer precision. We have demonstrated the use of such a system for studying the surface-enhanced Raman scattering of LC molecules with gold nanoparticles dispersed within the medium. Finally, we have also demonstrated probing of chemical inhomogeneity in the LC medium by using the optically trapped plasmonic particle as a probe of the Raman signal using the SERS effect.

## V. ACKNOWLEDGEMENTS

This work was supported by the branch contributions of the Institute for Complex Adaptive Matter (H.M.), the NSF grants DMR-0847782 (T.L., B.S., and I.I.S.) and DMR-0844115 (H.M. and I.I.S), and by the Office of Basic Energy Sciences of the US Department of Energy under Contract No. DE-AC36-08GO28308 with the National Renewable Energy Laboratory (J.v.d.L., P.J.A., and H.M.).

## FIGURES

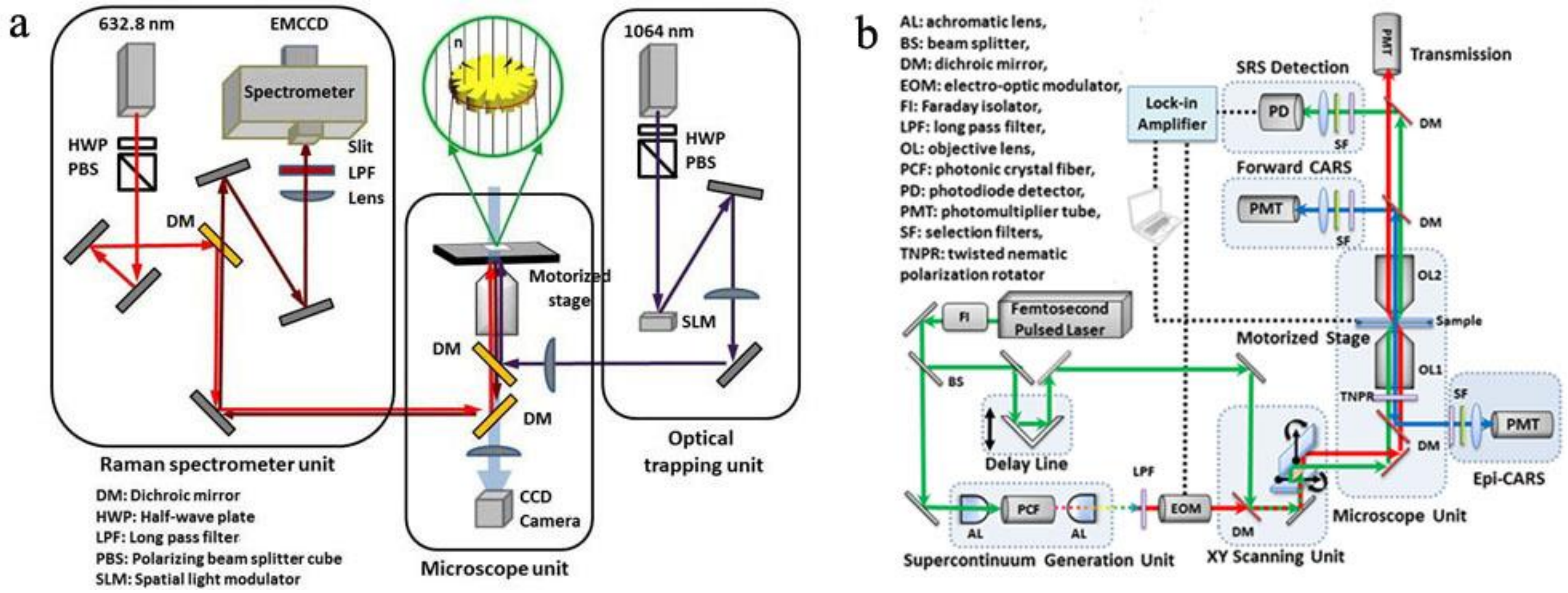

Fig. 1: a) A schematic diagram of an integrated system with homebuilt Raman micro-spectrometer and optical trapping. The inset shows the plasmonic nanoantenna in the form of the NB nanoparticle incorporated into LC sample. b) Schematic diagram of coherent Raman imaging setup with SRS and CARS microscopy modalities.

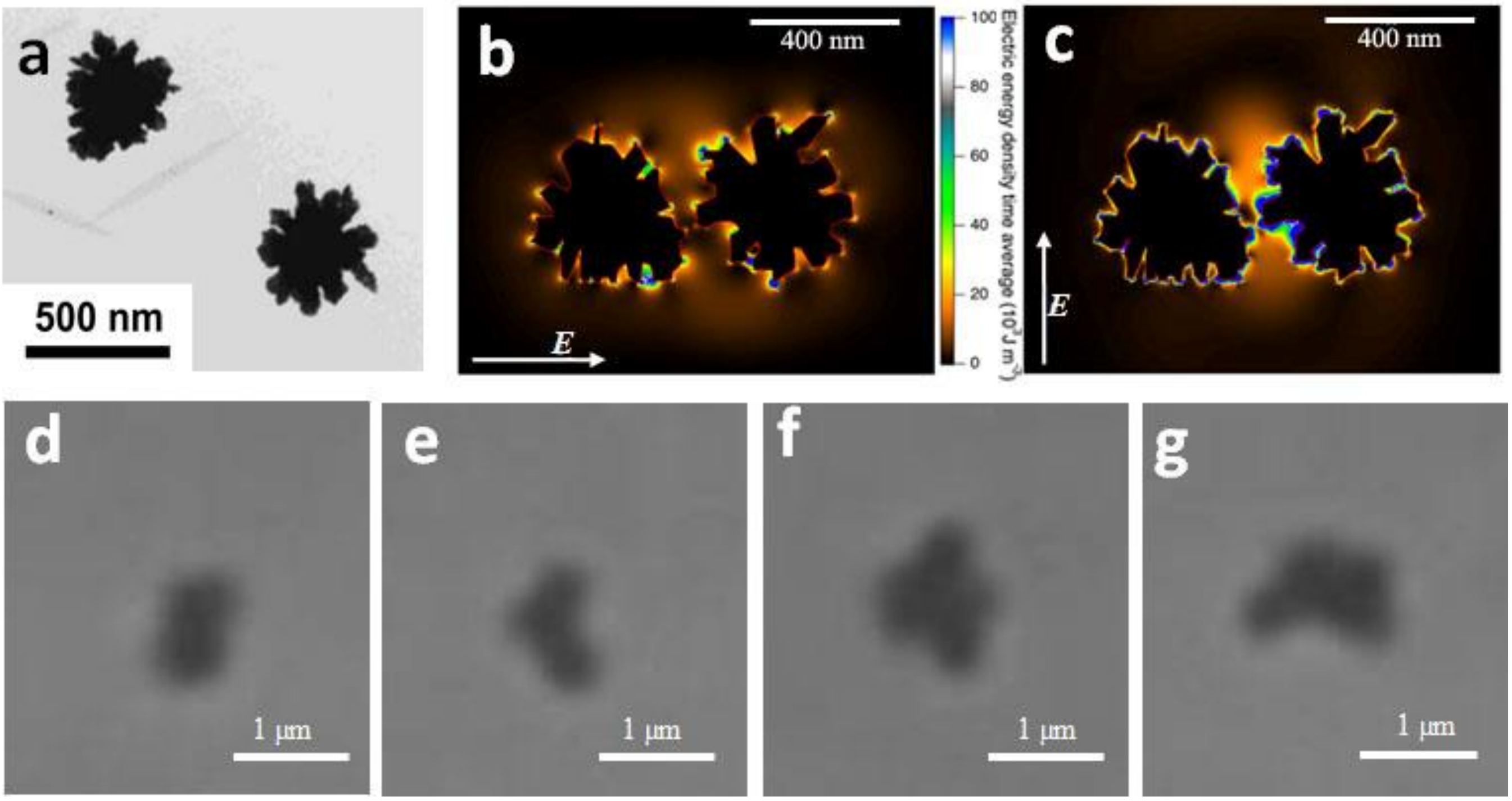

Fig. 2: (a) TEM image of the NB particles. (b,c) Simulations showing the electromagnetic field enhancement by NB particles in LC medium with linear polarization (electric field **E**) of the incident light. The spatial distributions of the electromagnetic field strength at the mid-plane of the NB nanoparticle (XY plane) oriented orthogonally to $\boldsymbol{n_0}$ were calculated and displayed (b,c) for two orthogonal directions of incident linearly polarized light (b) parallel and (c) perpendicular to the particles' center - to - center separation vector. Only the field inside the 5CB is shown for clarity. (d-g) Transmission-mode optical micrographs of (d) two, (e) three, (f) four, and (g) five NB particles assembled by use of optical trapping system in LC and held together by the LC elastic forces after the optically guided assembly.

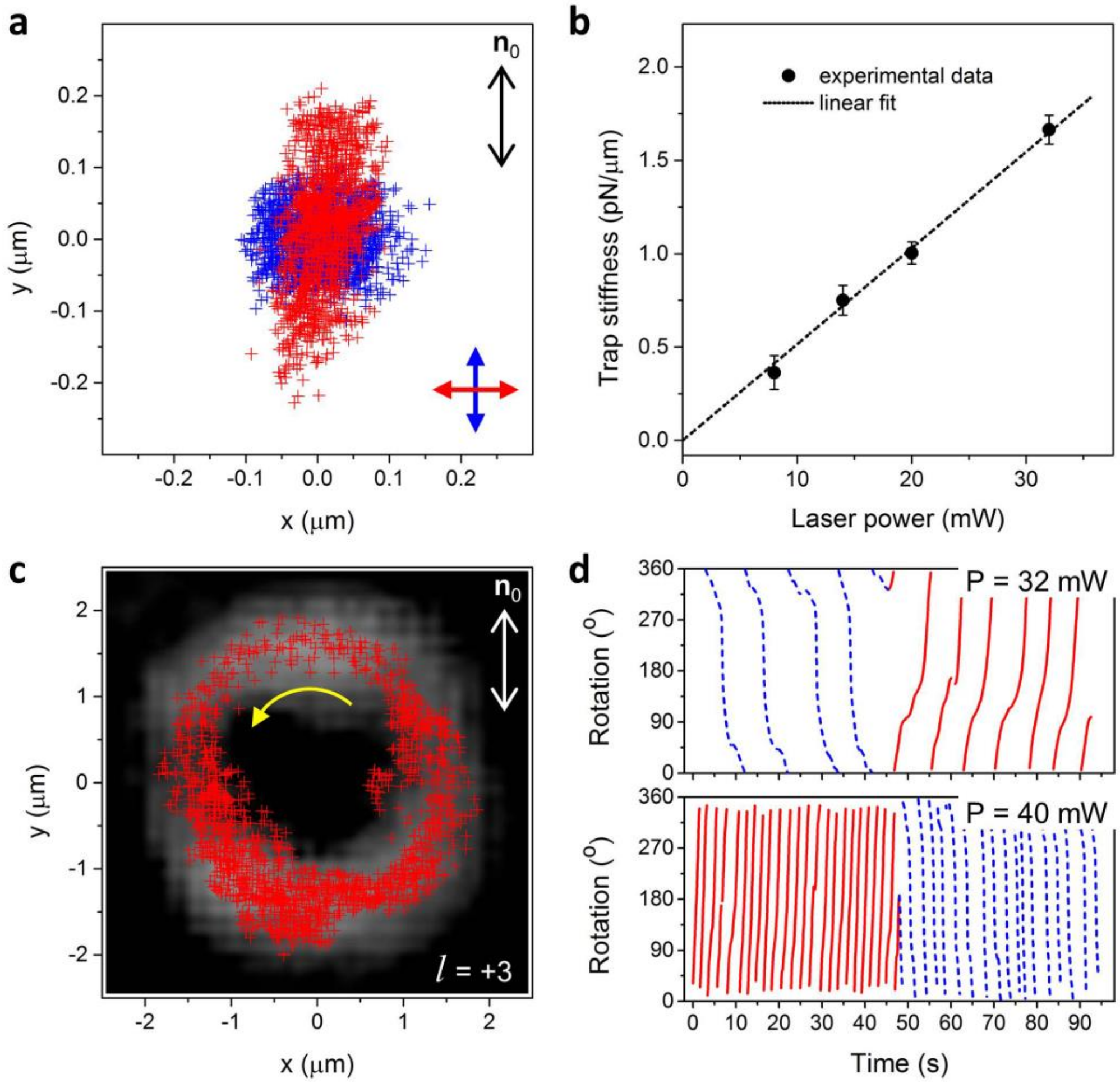

Fig. 3: Optical trapping of NB nanoparticles in nematic LC cells: (a) displacement of the trapped NB nanoparticle in a 10-µm-thick planar cell probed for two orthogonal polarizations (red and blue arrows) of the trapping beam at $P$ = 32 mW; (b) average trap stiffness vs laser power measured in a 3-µm-thick homeotropic cell; (c) counterclockwise rotational displacement (yellow arrow) of NB nanoparticle in the Laguerre–Gaussian mode of trapping beam with charge $l$=+3; (d) clockwise (red solid lines) and counterclockwise (blue dashed lines) rotation of the NB particle at laser power $P$ = 32 mW and $l$=±3 (top) and $P$ = 40 mW and $l$=±2 (bottom). Error bars in (b) correspond to the standard error of determination of trap stiffness.

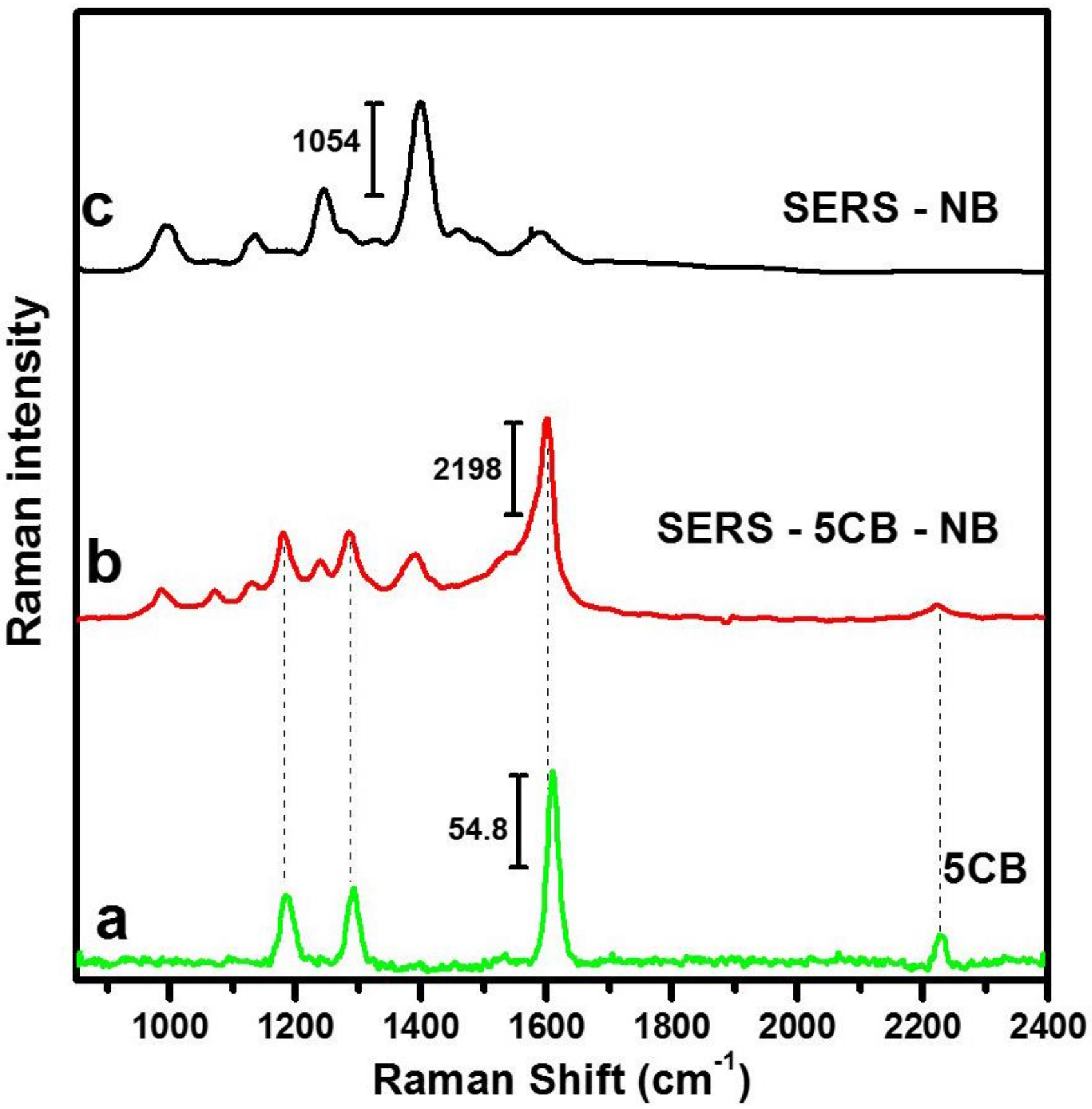

Fig. 4: Raman spectra of 5CB with incorporated NB particles collected from (a) the region without NB particles and (b) around a single NB nanoparticle. (c) Raman spectrum of the NB particles spin-coated on a glass substrate. Note the different scale in Raman intensity indicated by the vertical bars.

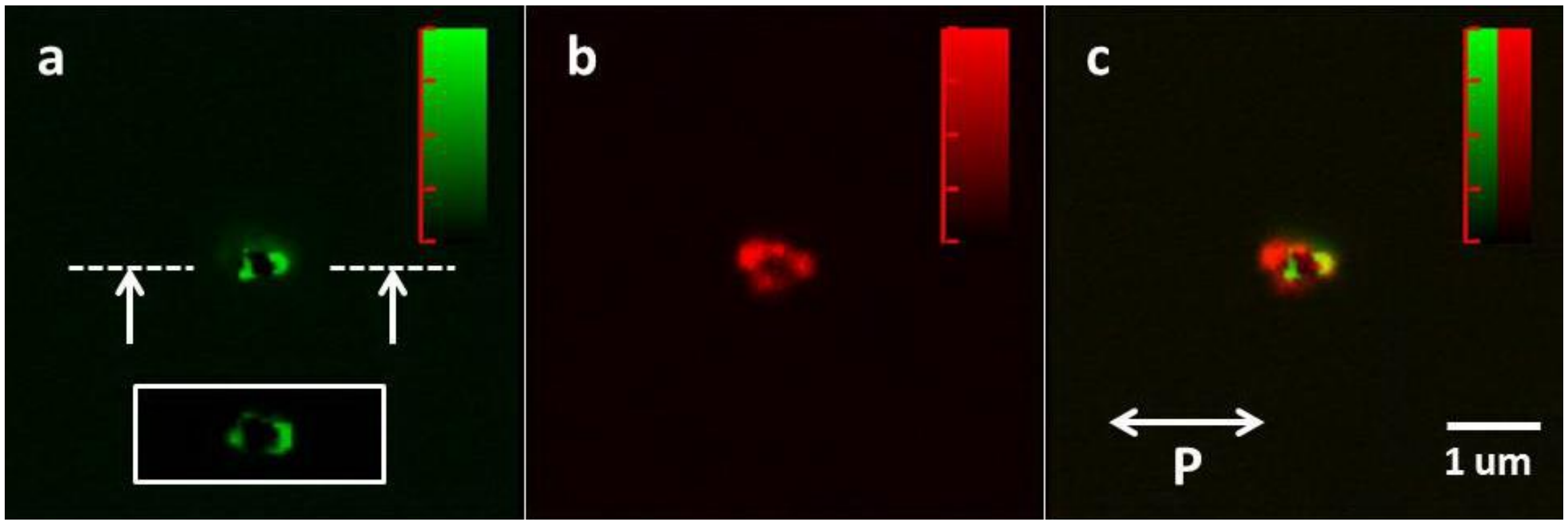

Fig. 5: Surface enhanced (a) SRS-PM and (b) CARS-PM images of homeotropically aligned 5CB samples with the signal enhancement around the dispersed NB particles. (c) Superimposed image of (a) and (b). Inset in (a) shows cross-sectional image along white dashed line.

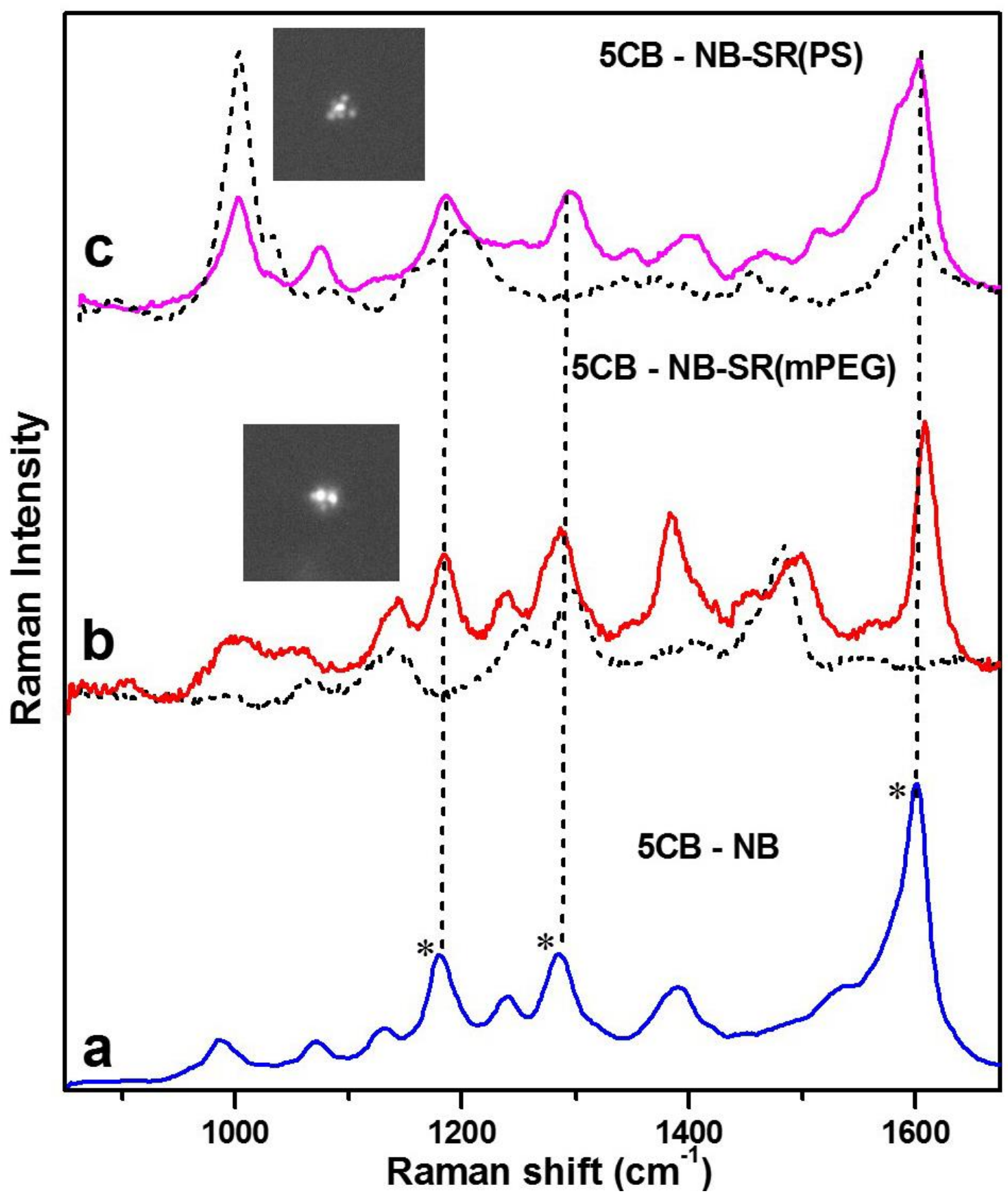

Fig. 6: Surface enhanced Raman spectra collected from (a) NB particles in 5CB and assemblies of NB and SR particles capped with (b) mPEG and (c) PS respectively. Dotted lines represent the Raman spectra corresponding to the capping ligands in (b-c). Star symbols represent the characteristic Raman lines of 5CB. The insets show the dark field micrographs of the NB-SR assemblies corresponding to the spectra shown in the respective panels.

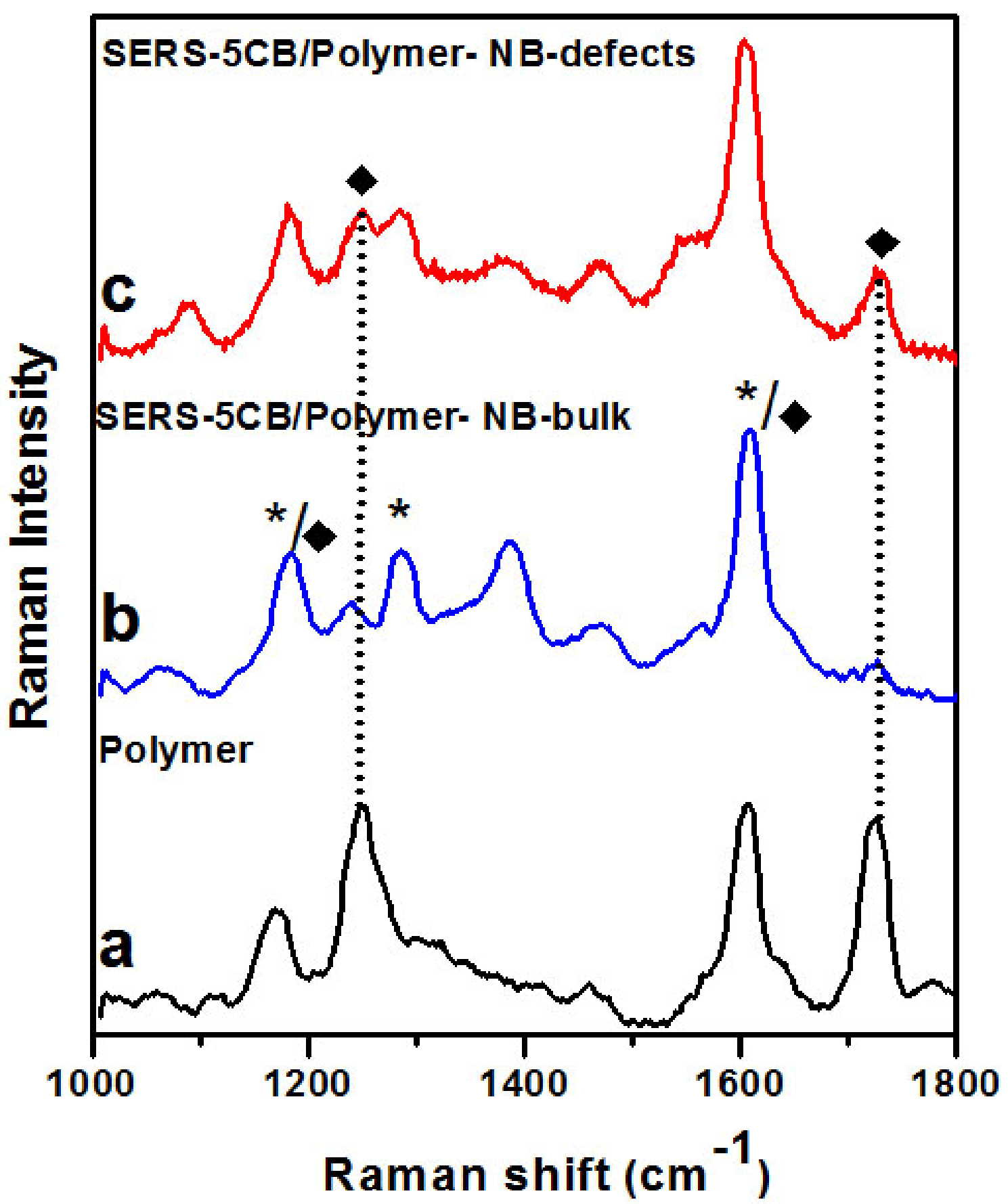

Fig. 7: Raman spectra collected from (a) polymer only and NB particles located (b) in the bulk and (c) at defects of the partially polymerized nematic LC sample. The star and diamond symbols represent the characteristic Raman lines from 5CB and polymer, respectively.